# Fast Dynamic Memory Integration in Co-Simulation Frameworks for Multiprocessor System on-Chip

O. Villa[1]   P. Schaumont[1]   I. Verbauwhede[1]   M. Monchiero[2]   G. Palermo[2]

[1] UCLA - {oreste, schaum, ingrid}@ee.ucla.edu
[2] Politecnico di Milano - {monchier, gpalermo}@elet.polimi.it

## Abstract

*In this paper is proposed a technique to integrate and simulate a dynamic memory in a multiprocessor framework based on C/C++/SystemC. Using host machine's memory management capabilities, dynamic data processing is supported without compromising speed and accuracy of the simulation. A first prototype in a shared memory context is presented.*

## 1. Introduction

MultiProcessor Systems on-Chip (MPSoC) [3] are an appealing solution for complex applications and are becoming feasible in contemporary technology. Especially for these complex systems, designers need frameworks to fast evaluate different possible implementations of the target architecture that covers interconnect, communications protocol, and topology is needed. Typical applications that run on MPSoC manage large amount of data (such as audio/video processing). Simulating these application, memory management capabilities are required. If these details are introduced in the model the memory's simulation becomes a significant portion of the workload for the simulator, affecting the performance in terms of simulation speed. Therefore a good memory model for exploration should: (I) allow the execution of complex applications with huge amounts of dynamic data (II) be accurate (III) have a low overhead (IV) be easy to design. The contribution of this paper is the development of flexible and efficient techniques based on the use of the host machine dynamic memory capabilities.

## 2. Framework

Today Co-Simulation frameworks for MPSoC are based on the integration of ISSs with simulation kernels. A typical simulation framework can be seen in Figure 2, where dashed boxes represent our contribution. In most of the traditional simulation frameworks, the memory subsystem is modeled as standard hardware modules. Unless complex and slow dynamic memory models are added, static memories implemented as tables are used. In our approach, dynamic hardware memory modules are simulated using a cycle true memory wrappers to access at the operating system functions which manages the host machine memory management capabilities. Dynamic memory operations like allocation, write/read and deallocation are implemented as communications between the hardware modules, or ISS, and the shared memory's wrapper. These operations are mapped by the wrappers in the native host machine functions, improving simulation speed. The wrapper guarantees the simulation accuracy using parameters of delays which can be dynamic and data dependent. Mechanisms to manage pointer arithmetic for user defined data-types, to model finite size memories and to reserve pointers in a shared memory context are proposed. High level APIs used by the ISSs are also provided using a C formalism. Multiple dynamic shared memories are considered and methods to manage general data structures are work in progress.

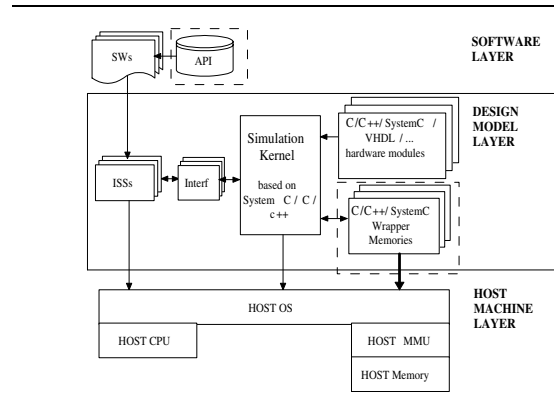

**Figure 1. Simulation Framework**

## 3. Dynamic Shared Memory

An implementation of our solution for multiple dynamic shared memories is employed in a system composed of sev-eral processing elements and an interconnect. In Figure 2, the overall simulated system is shown, as well as the details of the shared memory subsystem. Interconnect and shared memories are developed in C++. The simulation kernel is based on C++ [1] and the ISSs are ARM simulators [2].

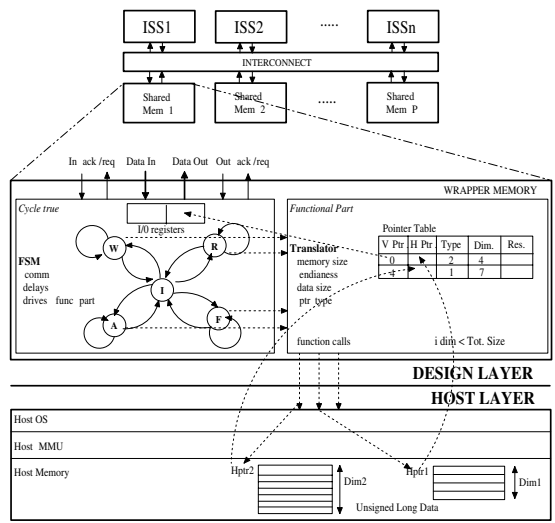

**Figure 2. Shared Memory Implementation**

The wrapper is composed of a cycle true part, described as a Finite State Machine (FSM), and a functional part, composed of a pointer table and a translator. In the FSM the communication is performed following a handshake protocol and incoming signals are evaluated cycle by cycle. Operations as *allocation*, *read/write* and *deallocation* are identified by an input code (*opcode*). The opcode and the shared memory address (*sm_addr*), which identifies the memory module, are the first data of every transaction between the ISSs and the wrapper. Endianess, data type translation and host machine functional calls are performed by the translator led by the FSM. Virtual pointers for the simulated architecture, real host pointers to retrieve the data in the host machine are stored in the pointer table, as well as the size of the allocated spaces, type of allocated spaces and a reservation bit used as semaphore to reserve the pointers. When an *allocation* operation is required, size (*dim*) and data type (*type*) must be sent. The FSM maps this communication in a *calloc(dim, DATA_SIZE)* operation to the host machine using the translator. A pointer is then returned by the host machine and stored in the table with the *type* and *dim* information. A virtual pointer (*Vptr*) is sent back to the ISS completing the transaction. Every new allocation is a new entry in the pointers table and a finite size memory can be simulated denying other allocations when the sum of the dimension reaches a prefixed limit. When a *write* operation is required, the virtual pointer (*Vptr*) and the data (*DataIn*) are provided. The pointer table is checked in order to see if the *Vptr* is valid and to retrieve the real host machine's pointer (*Hptr*). The *DataIn* is stored in the *Hptr* host location. Similar mechanisms are adopted for a *read* operation. In this case, the data retrieved in the host memory is sent as output data (*DataOut*) to the communication port by the FSM. For a *free* operation (*deallocation*), it is necessary to provide the *Vptr*. Then the entry associated to the Vptr in the pointer table is removed, the table is re-compacted and the dimension of the deallocated memory is subtracted to the total memory size. Finally, a *free* function, with the correspondent *Hptr* as argument, is performed on the host. The generation of the *Vptr* is performed in such way that every new *Vptr* is obtained summing the value of the previous *Vptr* in the table with the size of the previous allocated space (see the table in Figure 2). The first *Vptr*'s value is zero by default. If pointer arithmetic is used to access to the memory, incoming virtual pointers values that are not present in the table are received. In this case, the true host memory space is retrieved checking in the pointer table which space belongs to the *Vptr*. The *Hptr* is, therefore, calculated adding an appropriate offset. When indexed structures are exchanged, the dimension of the transmitted arrays must be also sent together with *Vptr*. I/O registers are substituted by I/O arrays, where the incoming and outgoing data are temporally stored until the communication is completed. Afterward, they are moved to the host machine, employing a similar mechanism used for scalar data. Data coherence is guaranteed in our design preserving the *Vptr* by means of the reservation bit which is set by an ISS that want to protect the pointer. To model data dependent latencies, a set of *delay* parameters can be used in the FSM. Multiple instances are easily managed, since the host machine provides the generation of a different host pointer for every allocation. High level APIs very similar to the host machine functions are used by the ISSs. Different hardware devices that might be connected on the system can access to the memories using low level communication.

## 4. Conclusions

Preliminary experimental results, obtained simulating the GSM algorithm, show that the overhead in terms of simulation speed introduced by this model is very low. Comparing the simulation speed of 4 ISSs with one memory and interconnect and this of 4 ISSs with interconnect and 4 memories we found a degradation of simulation speed of 20%.

## References

[1] *GEZEL*. www.ee.ucla.edu/∼schaum/gezel.
[2] *SimIT-ARM*. www.princeton.edu/∼wquin/armsim.htm.
[3] W. Wolf and A. Jerraya. *Multiprocessor System on-Chips*. Morgan Kaufman Publishers, 2004.